\documentclass[conference,10pt]{IEEEtran}
\usepackage{amssymb,amsmath,epsfig,psfrag,epstopdf,enumerate}
\usepackage[english]{babel}
\usepackage{color}
\usepackage{graphicx}

\usepackage[normalem]{ulem}
\usepackage{varioref}
\usepackage{balance}
\usepackage{subfig}

\begin{document}

% ------ % Title. % ------
\title{Golden Angle Modulation:  \\
Approaching the AWGN Capacity}

% ------------------- % Single address. % -------------------

\author{
\IEEEauthorblockN{Peter Larsson \emph{Student
Member, IEEE},  Lars K. Rasmussen \emph{Senior
Member, IEEE},\\ Mikael Skoglund, \emph{Senior
Member, IEEE}}%
\thanks{The authors are with the ACCESS Linnaeus Center and the School of
Electrical Engineering at  KTH Royal Institute of  Technology, SE-100 44
Stockholm, Sweden.}
}
\maketitle

%%%%%%%%%%%%%%%%%%%%%%%%%%%%%%%%%%%%%%%%%%%%%%%%%%%%%%%%%%%%%%%%%%%%%%%%%%%%%%%%%%
\begin{abstract}
In this work, targeting, e.g., future generation cellular, microwave-links, or optical fiber systems, we propose a new geometric shaping design for golden angle modulation (GAM) based on a (double) truncated Gaussian input distribution. The design improves the mutual information (MI), and the peak-to-average power ratio, over the full signal-to-noise ratio (SNR) range relative to two key GAM schemes introduced in \cite{Larsson17a,Larsson17b}. Inspired by the proposed geometric shaping, a simpler, SNR-dependent, design is also suggested. The performance is numerically evaluated with respect to MI and compared with classical modulation schemes. With the proposed design, the SNR can be decreased relative to classical quadrature amplitude modulation, even for relatively modest target spectral efficiencies. As the GAM design can approach the Gaussian channel capacity, the power/energy efficiency is expected to improve.
\end{abstract}


\section{Introduction}
\IEEEPARstart{F}{uture} radio/optical communications systems require enhanced communication rates, and/or reduced peak/average transmit power/energy. In this respect, the modulation scheme is important. It is desirable that the average mutual information (MI) vs. the signal-to-noise ratio (SNR), $S$, of the modulation schemes, is as close to the additive white Gaussian noise (AWGN) (Shannon) capacity, $C=\log_2(1+S)$ [b/Hz/s] as possible. Well known modulation signal constellations are quadrature amplitude modulation (QAM), phase shift keying (PSK), star-QAM \cite{HanzoNgKellWebb04}, and amplitude PSK (APSK) \cite{ThomasWeidDurr74}.
QAM, the most studied and deployed of all constellation designs, exhibits an asymptotic 1.53 dB SNR-gap (a shaping-loss) between the MI and the AWGN capacity \cite{ForneyUnger98}. Hence, QAM waste $\approx 30$\% more transmit power than needed at high rates.
Geometric- and probabilistic-shaping have been proposed to overcome this shaping-loss.  Some works on geometric-shaping are on asymmetric constellations in \cite{DivsalarSimonYuen87}, on nonuniform-QAM in \cite{BettsCaldeLaroi94}, on nonuniform-PAM for AWGN in \cite{SommerFett00}, and on PSK/PAM capacity optimization in \cite{BarsoumJoneFitz07}. Similarly, some works on probabilistic shaping are as a survey in \cite{ForneyGallLangLongQure84}, on $N$-dimensional sphere/cube constellations in \cite{CalderbankOzar90}, on Maxwell-Boltzmann distribution based probabilistic shaping in \cite{KschischangPasu93}, and on APSK in   \cite{LiuXiePengYang11,XiangVale13,Meric15}.
While geometric shaping can be applied to QAM, a constellation with circular symmetry (like APSK) is often preferred since the capacity achieving (complex Gaussian) distribution is circular symmetric. Yet, one issue with geometrically-shaped APSK is that the number of constellation points in each ring depends on the shaping.

Recently in \cite{Larsson17a}, and extended in \cite{Larsson17b}, the \textit{Golden angle modulation} (GAM) framework was introduced. There are two key design features in GAM. The first is separating consecutively indexed constellation points with the golden angle in the phase domain. The second is allowing the radial and the phase distribution of constellation points to have different forms and dependency on the constellation point index.
More formally, the GAM signal constellation points can be expressed as
\begin{align}
x_m&=r_m\mathrm{e}^{i\phi_m}, \, m=\{1,2,\ldots,M\},
\label{eq:Eqxm}
\end{align}
where $\phi_m=2\pi \varphi m$, $\varphi\triangleq (3-\sqrt{5})/{2}\approx 0.381$, and $2\pi \varphi$ is the golden angle in radians.
For large enough constellations, this discrete, spiral-related, design offers a near ideal circular symmetric design, a near ideal uniform phase distribution, and allows for geometric radial shaping.
More specifically in \cite{Larsson17a,Larsson17b}, two key GAM-designs where proposed. The first one, geometric bell-shaped GAM (GB-GAM), had magnitudes
\begin{align}
r_m&=c_\textrm{gb}\sqrt{\ln{\left(\frac{M}{M+1-m}\right)}},  \, m\in\{1,2,\ldots,M\},
\label{eq:Eqgb}
\end{align}
where  $c_\textrm{gb}\triangleq\sqrt{{ \bar P}/{(\ln{M}-\ln(M!)/M)}}$, and average power $\bar P$. The second one,  disc-shaped GAM (disc-GAM), had
\begin{align}
r_m&=c_\textrm{disc}\sqrt{m}, \, m\in\{1,2,\ldots,M\},
\label{eq:Eqdisc}
\end{align}
where $c_\textrm{disc}\triangleq\sqrt{{2\bar P}/{(M+1)}}$.

We note that relatively few works, preceding GAM, consider spiral-based designs. Analog spiral-based modulation was proposed in \cite{ThomasMayWelt75,KvecherReph06,KravtsovRaph07}. The first work uses interconnected semicircular segments of increasing radius, whereas the latter two studied Archimedean spirals with amplitudes $f(x)=x\mathrm{e}^{i x}, \, x\geq 0,$ in \cite{KvecherReph06}, and $f(x) = g(x)\mathrm{e}^{i g(x)}, \, g(x)\geq 0$, in \cite{KravtsovRaph07}. In \cite{JafargholiEsmiMousHoss06,KwakSongParkKwon08}, discrete spiral-based modulation schemes were proposed. Whereas the former examined a logarithmic spiral design, the latter proposed a design with four intertwinned Archimedean spirals. In the related area of joint source-channel coding, \cite{HeklandOienRams05, FloorRams06, ZaidiKhorSkog09}, also considered Archimedean spiral-based mapping.
The GAM framework does not adopt, nor limit itself to, the Archimedean spiral design. This is not just because the Archimedean spiral has been used several times before, but, more importantly, since it by design, i.e. not decoupling phase and magnitude forms, defy the notion of radial shaping with near-ideal uniform phase distribution. Such characteristic, as offered by GAM, is desirable, e.g. for AWGN capacity achieving, and/or a PAPR-controlled, designs. In addition, none of the prior works have used, nor benefited from, the golden angle-based phase arrangement of constellation points.

In \cite{Larsson17a,Larsson17b}, it was seen that the MI of GB-GAM was close to the channel capacity for, say, $0\leq MI\approx H/2$, where $H=\log_2(M)$ is the signal constellation entropy. Disc-GAM, on the other hand, performed best when $MI\lessapprox H$. Hence, there is a wide SNR-range, where neither GB-GAM, nor disc-GAM, perform optimally. An effort to handle this in \cite{Larsson17a,Larsson17b}, was by optimizing the magnitudes $r_m$ to maximize the MI for every SNR. This problem could not be solved analytically, and a numerical optimization solving approach could only handle $M\lessapprox16$ constellation points. Hence, it is of interest to find an analytical expression for the magnitudes $r_m$ that improves the MI for the full SNR-range, $S\in (0,\infty)$, relative to GB-GAM, disc-GAM, and classical signal constellations. It is also of interest to parameterize the analytical $r_m$ expression in one, or two, variables that allows for numerical optimization of larger constellations sizes.

\subsection{Truncated Geometric Bell-shaped GAM}
\label{sec:Sec5p5d2d2d0}

It can be recognized that GB-GAM and disc-GAM can be seen as resulting from two extremes of a truncated Gaussian input distribution. GB-GAM would then approximate a (non-truncated) Gaussian pdf, and disc-GAM would approximate a truncated Gaussian pdf when the radius approaches zero. We can approach this line of reasoning more formally. The overall procedure is to first find a continuous input distribution that (closely) approximates the MI-maximizing distribution, and then using this distribution together with the GAM framework and inverse sampling to determine constellation point magnitudes $r_m$.
First, the MI between a continuous input signal $X$, and a continuous output signal $Y$, can be lower bounded as follows,
$I(Y;X)
=h(Y)-h(Y|X)
=h(Y)-h(W)
\geq h(X)-h(W)$,
where the fact that conditioning reduces entropy, $h(Y)\leq h(Y|W)=h(X)$, is used in the last step. Hence, instead of finding an input distribution for $X$ that maximize the entropy $h(Y)$, the  entropy $h(X)$ is maximized, thereby giving a lower bound to $I(Y;X)$.

Since GAM is near circular symmetric, i.e. for a sufficiently large $M$, we target a continuous circular symmetric input distribution $f(x_\textrm{re},x_\textrm{im})$ that maximizes $h(X)$ for an average power constraint. For notation simplicity and clearness below, we use the notation $f(u,v)$ instead of $f(x_\textrm{re},x_\textrm{im})$. Moreover, for generality, and control over the PAPR, we let the circular symmetric density $f(u,v)$ be nonzero for magnitudes less than an outer radius, $\rho_\textrm{o}$, but also greater than an inner radius,  $\rho_\textrm{i}$. More precisely, the pdf is assumed to fulfill

\begin{equation}
\begin{aligned}
f(u,v) &\geq 0,  \, (u,v)\in \mathcal{A}, \\
f(u,v) &= 0,  \,  \text{Otherwise},
\end{aligned}
\end{equation}
where
$\mathcal{A}\in \{u,v:\rho_\textrm{i}^2\leq u^2+v^2 \leq \rho_\textrm{o}^2\}$.

The optimization problem can now be formulated as
\begin{equation}
\begin{aligned}
& \underset{f(u,v)}{\text{maximize}}
& &-\iint_{\mathcal{A}} f(u,v)\ln\left(f(u,v)\right) \, \mathrm{d}u \, \mathrm{d}v, \\
& \text{subject to}
& & \iint_{\mathcal{A}} (u^2+v^2)f(u,v)  \, \mathrm{d}u \, \mathrm{d}v=1, \\
& & & \iint_{\mathcal{A}} f(u,v)  \, \mathrm{d}u \, \mathrm{d}v=1.
\end{aligned}
\end{equation}
This problem can, due to convexity of the entropy, be solved by classical Lagrangian optimization. The solution is straightforward and is simply a truncated 2-dim Gaussian pdf. Writing it on the standard bi-variate Gaussian form, with an undetermined normalization constant $c_1$, we have $f(u,v)=\frac{c_1}{\pi \sigma^2}\mathrm{e}^{-(u^2+v^2)/\sigma^2}$, where $(u,v)\in \mathcal{A}$. Since $\rho_\textrm{i}$ and $\rho_\textrm{o}$ can be tuned to any value, without loss of generality, we let $\sigma^2=1$. As the discrete GAM magnitudes $r_m$ are sought after, it useful to transform $f(u,v)$ to a marginal density $f(\rho)$ with respect to a magnitude $\rho$. Via Euclidean-to-polar coordinate variable substitution, $u=\rho \sin(\phi)$ and $v=\rho \cos(\phi)$, and integrating over a uniform distribution in phase, we get

\begin{equation}
f(\rho)=
\left\{
\begin{matrix}
\frac{c_1}{\pi}\mathrm{e}^{-\rho^2}2\pi \rho
,&  \rho_\textrm{i}\leq \rho\leq \rho_\textrm{o},\\
0
,& \text{Otherwise}.\\
\end{matrix} \right.
\end{equation}

Integrating the pdf,  $\int_{\rho_\textrm{i}}^{\rho_\textrm{o}} f(r) \, \mathrm{d}r=1$, yields the constant $c_1=1/(\mathrm{e}^{-\rho_\textrm{i}^2}-\mathrm{e}^{-\rho_\textrm{o}^2})$. The corresponding cdf is then
\begin{align}
F(\rho)&=\frac{\mathrm{e}^{-\rho_\textrm{i}^2}-\mathrm{e}^{-\rho^2}}
{\mathrm{e}^{-\rho_\textrm{i}^2}-\mathrm{e}^{-\rho_\textrm{o}^2}},
&  \rho_\textrm{i}\leq \rho\leq \rho_\textrm{o}.
\end{align}

\begin{figure*}[htp!]%
 \vspace{-0.4cm}
\centering
\hspace{-10pt}%
\subfloat[][]{%
\label{fig:ConstG2-b}%
 \includegraphics[width=6.25cm]{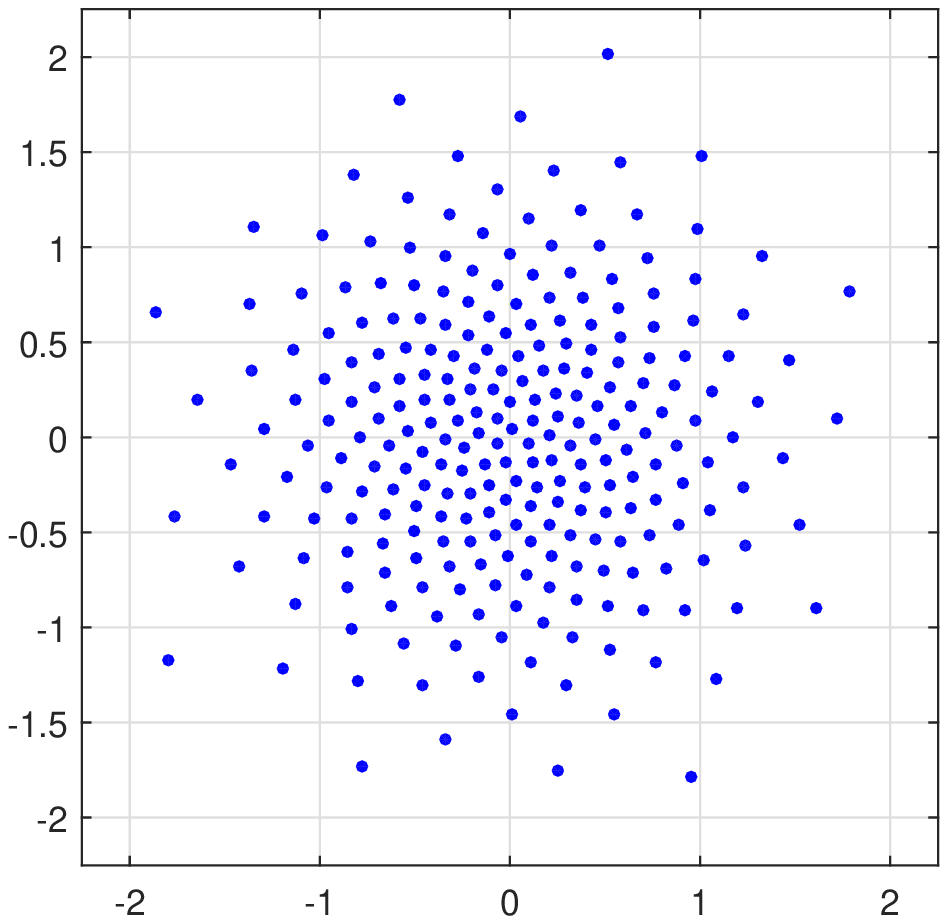}}
\hspace{-10pt}%
\subfloat[][]{%
\label{fig:ConstG2-c}%
 \includegraphics[width=6.25cm]{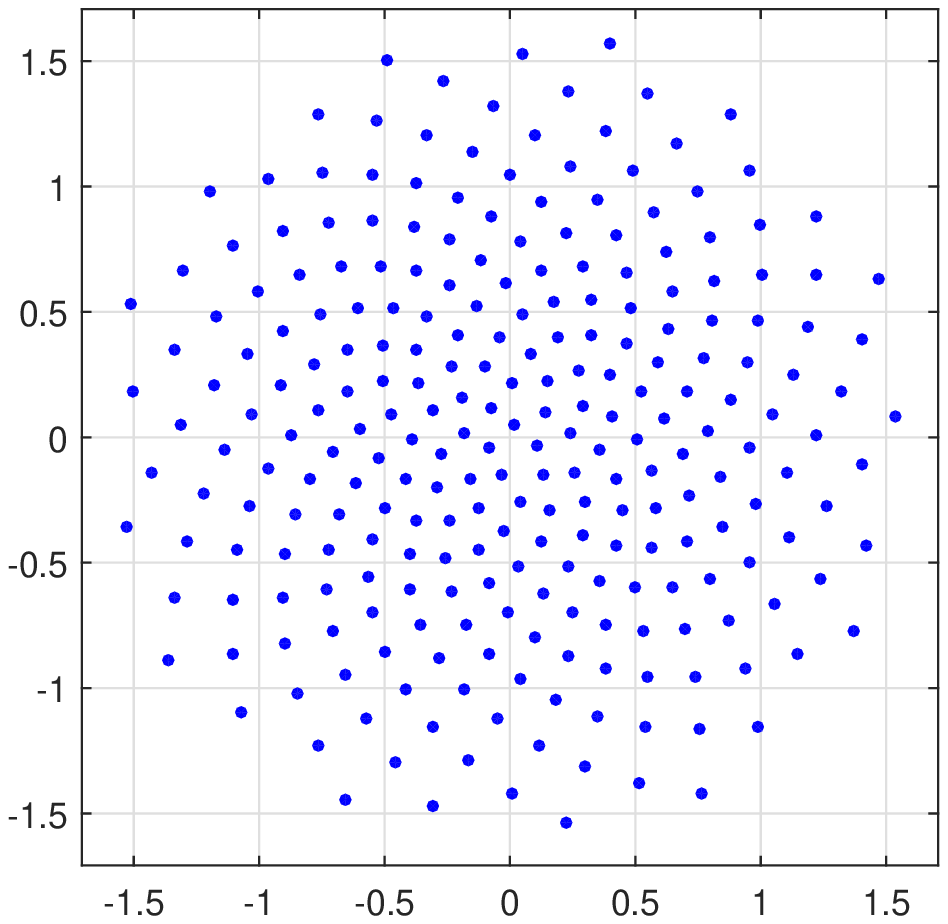}}%
\hspace{-10pt}%
\subfloat[][]{%
\label{fig:ConstG2-d}%
 \includegraphics[width=6.25cm]{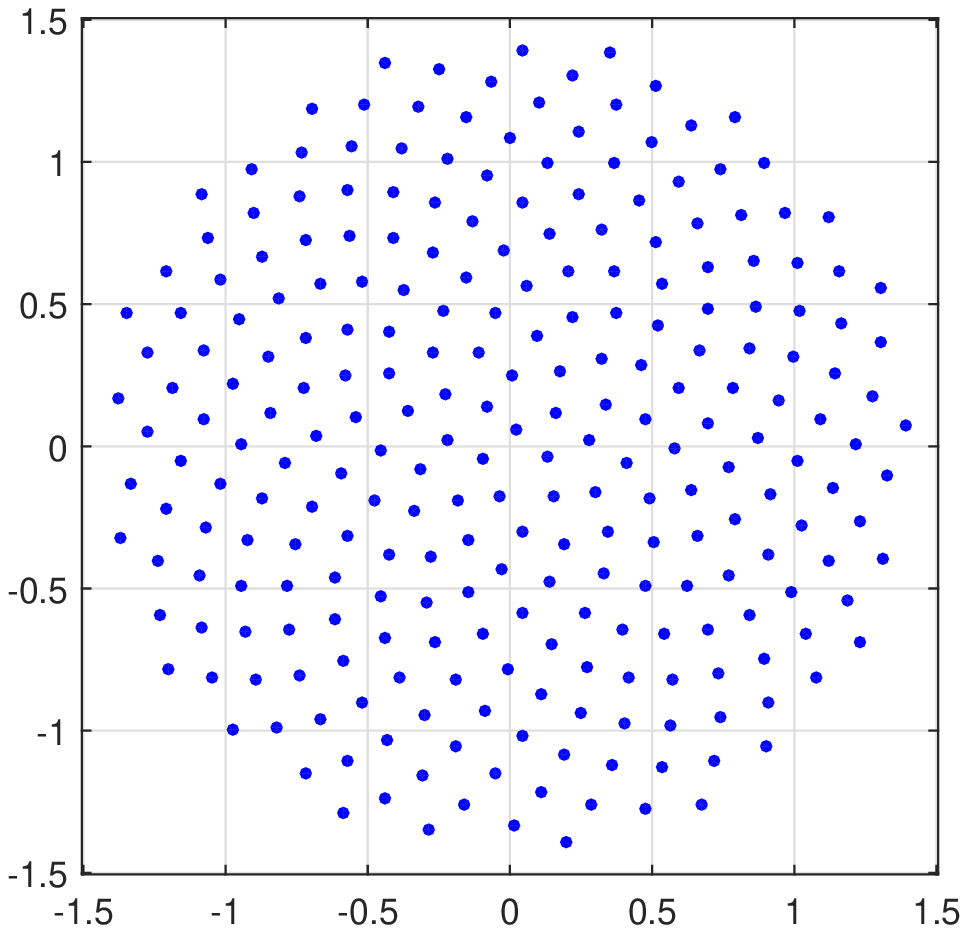}}%
\caption[ .]{Signal constellations of TGB-GAM, $\rho_\textrm{i}=0$, $M=256$:
\subref{fig:ConstG2-b} $S= 10$ dB;
\subref{fig:ConstG2-c} $S= 22.5$ dB;
\subref{fig:ConstG2-d} $S=35$ dB.}%
\label{fig:ConstG2}%
 \vspace{-0.35cm}
\end{figure*}

\begin{figure*}[htp!]%
\centering
\hspace{-10pt}%
\subfloat[][]{%
\label{fig:MagnG2-b}%
 \includegraphics[width=6.25cm]{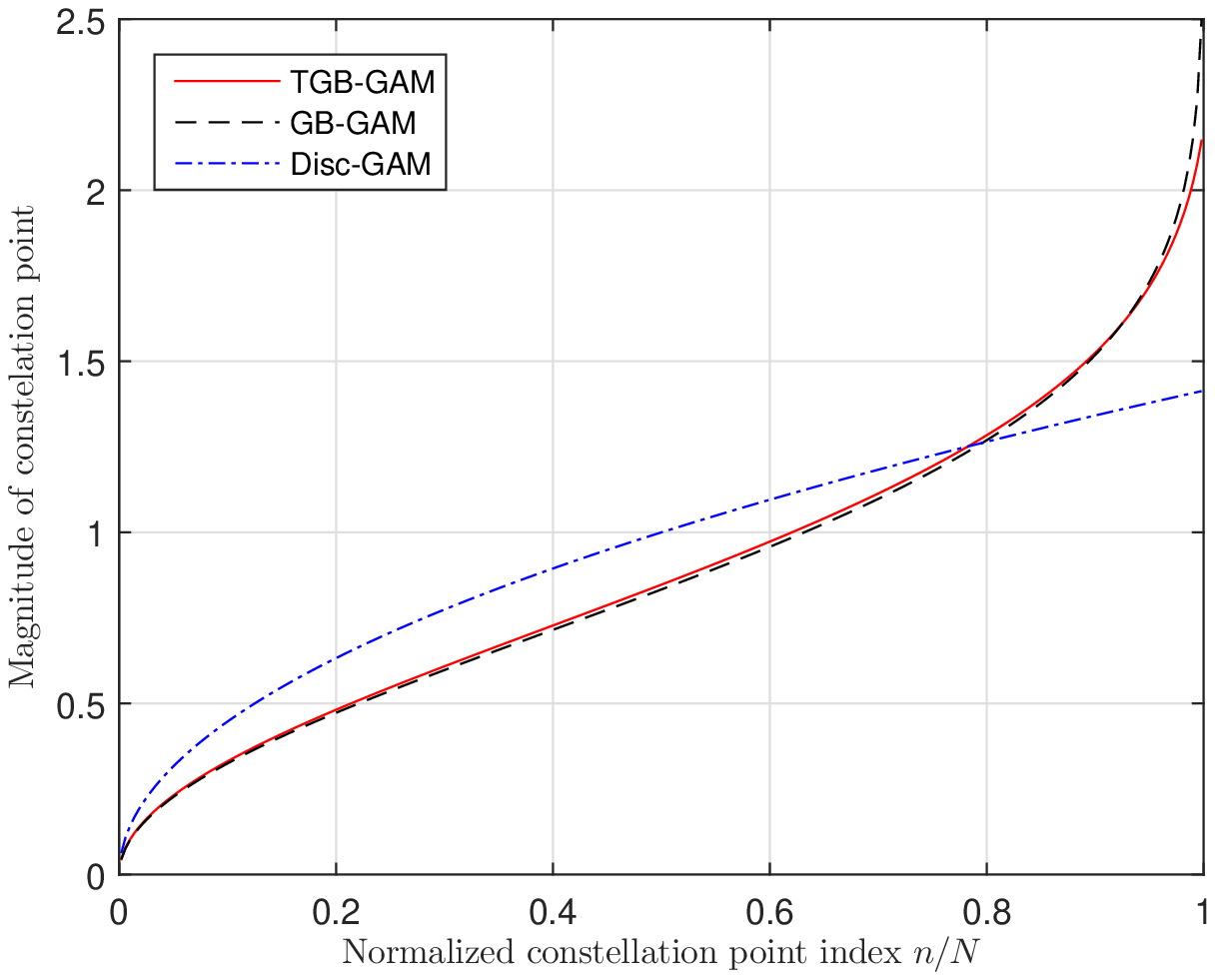}}
\hspace{-10pt}%
\subfloat[][]{%
\label{fig:MagnG2-c}%
 \includegraphics[width=6.25cm]{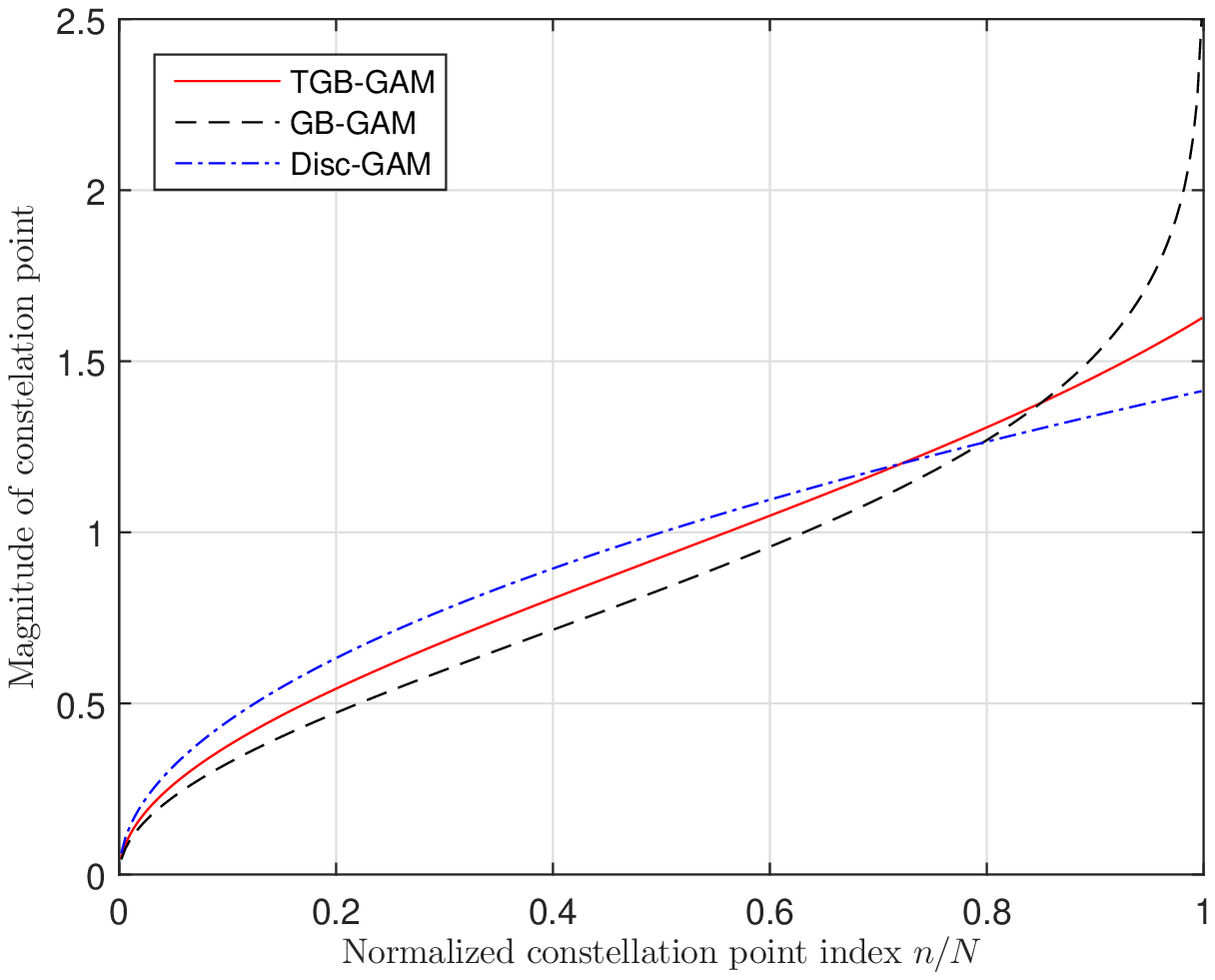}}%
\hspace{-10pt}%
\subfloat[][]{%
\label{fig:MagnG2-d}%
 \includegraphics[width=6.25cm]{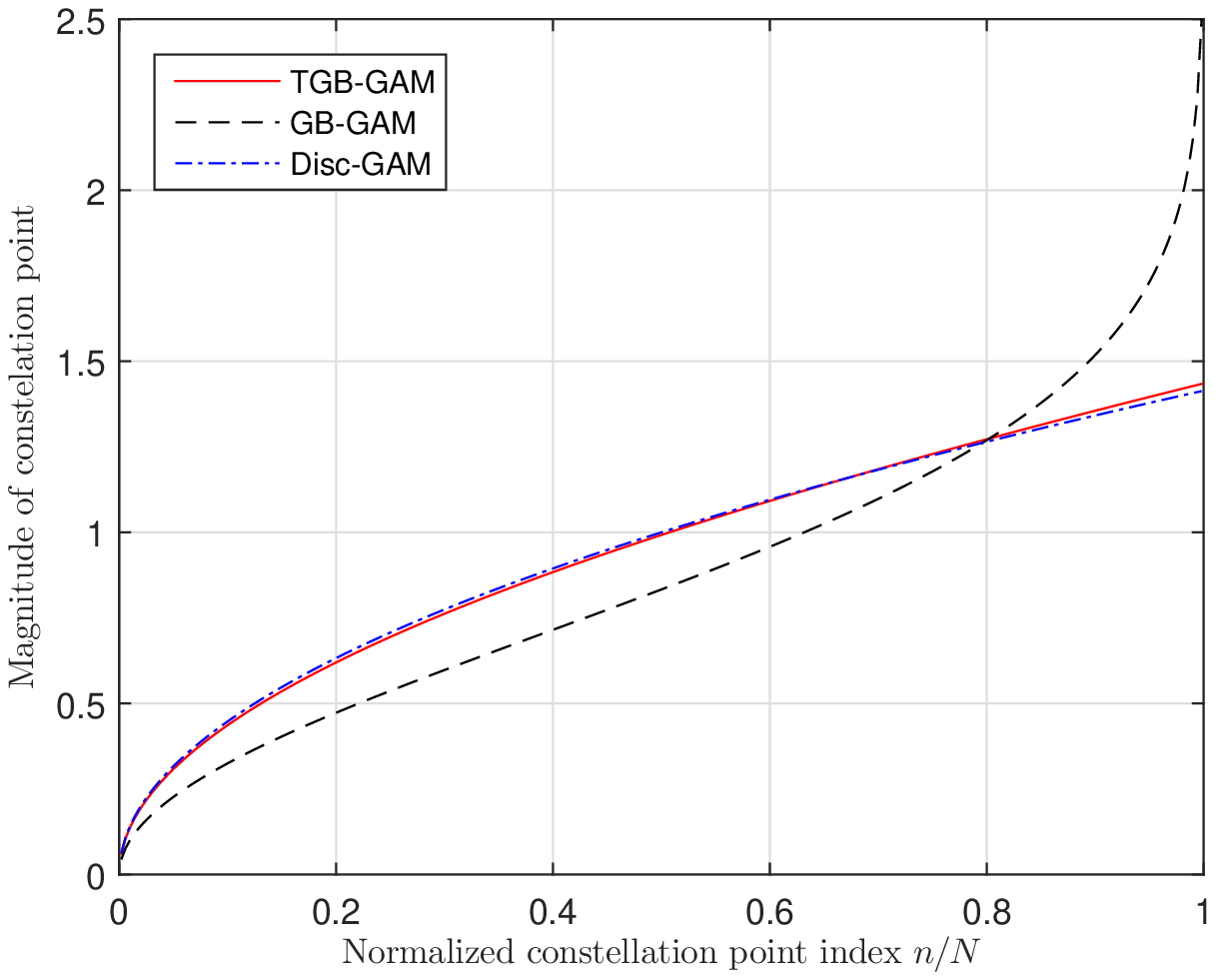}}%
\caption[ .]{Magnitude distribution of TGB-GAM, $\rho_\textrm{i}=0$, $M=256$:
\subref{fig:MagnG2-b} $S=10$ dB;
\subref{fig:MagnG2-c} $S=22.5$ dB;
\subref{fig:MagnG2-d} $S=35$ dB.}%
\label{fig:MagnG2}%
 \vspace{-0.35cm}
\end{figure*}

Similar to this work, in \cite{Smith71,ShamaiBD95,HosoyaYash16}, a peak-power constrained input distribution was considered. Specifically, the complex AWGN channel was studied in \cite{ShamaiBD95}. It was found that the optimal input distribution is discrete in the radial domain, but uniformly continuous in phase. As a sub-optimal input distribution, they derived (based on the same line of reasoning as above) a peak-power truncated complex Gaussian distribution. However, no actual modulation scheme was proposed. Moreover, the result on continuous uniformly distributed phase is incompatible with the idea of discrete signal constellation points.
Even if a modulation scheme had been proposed, say APSK-based, it is not obvious how to combine APSK with the truncated complex Gaussian input distribution assumption in a simple and well-performing manner. The designer of an APSK scheme must carefully split the total number of constellation points among the rings, and arrange the distance among rings. The constellations must be designed, and optimized, for every distinct $M$. Certain values of $M$, e.g. prime numbered $M$, give APSK constellations with asymmetries.
GAM, on the other hand, together with the method of inverse sampling, is versatile enough to seamlessly adapt to a desired truncated complex Gaussian input distribution and any integer $M$. To more flexibly control PAPR, as said, we also let $0\leq \rho_\textrm{i} < \rho_\textrm{o}$.

To design a discrete constellation, we consider the method of inverse sampling, together with the GAM design, and the (double) truncated bi-variate Gaussian target  density. In inverse sampling, one equates the cdf $F(\rho)$ with a continuous uniformly distributed r.v., say $t\in(0,1)$, and solve for $\rho$. Generating $t$ uniformly on $(0,1)$, then gives $\rho$ with pdf $f(\rho)$. With large number of constellation points, we can approximate the uniformly distributed continuous r.v with a uniformly distributed discrete r.v. $\tau$ with a pmf with, e.g., $\delta(\tau=m/M)=1/M, m\in\{1,2,\ldots,M\}$, or alternatively with $\delta(\tau=(m-1/2)/M)=1/M$. Thus, we have $F(\rho_m)=m/M$, from which we solve for $\rho_m$ and get
\begin{align}
\rho_m
&=\sqrt{-\ln
\left(\mathrm{e}^{-\rho_\textrm{i}^2}-\frac{m}{M}\left(\mathrm{e}^{-\rho_\textrm{i}^2}-\mathrm{e}^{-\rho_\textrm{o}^2}\right)\right)}.
\end{align}

It is convenient to normalize the magnitudes $\rho_m$, to an average power constraint, $\bar P$, by letting $r_m=c_\textrm{tgb} \rho_m$, and where the constant $c_\textrm{tgb}$ is determined as below
\begin{align}
\frac{1}{M}\sum_{m=1}^M r_m^2=\bar P
\Rightarrow c_\textrm{tgb}=\sqrt{\frac{M\bar P}{\sum_{m=1}^M \rho_m^2}}.
\end{align}
Henceforth, we let $\bar P=1$, and the AWGN noise variance is then $\sigma^2=1/S$. In total, the (unit-power normalized) GAM constellation point magnitudes are
\begin{align}
r_m
&=\sqrt{\frac
{\ln
\left(\mathrm{e}^{-\rho_\textrm{i}^2}-\frac{m}{M}\left(\mathrm{e}^{-\rho_\textrm{i}^2}-\mathrm{e}^{-\rho_\textrm{o}^2}\right)\right)}
{ \frac{1}{M}\sum_{m'=1}^M \ln
\left(\mathrm{e}^{-\rho_\textrm{i}^2}-\frac{m'}{M}\left(\mathrm{e}^{-\rho_\textrm{i}^2}-\mathrm{e}^{-\rho_\textrm{o}^2}\right)\right)}}.
\label{eq:Eqrm}
\end{align}
Note that since the constellation is unit-power normalized, the resulting PAPR is simply $PAPR_\textrm{dB}=20 \log10(r_M)$ dB.

As $r_m$ depends only on two parameters  $\rho_\textrm{i}$ and $\rho_\textrm{o}$, a simplified optimization problem can be formulated compared to optimizing all $M$ values for $r_m$. The advantage of this is that numerical optimization is not seriously limited by constellation size. The optimization problem, including an optional PAPR constraint $PAPR_0$, can then be written
\begin{equation}
\begin{aligned}
& \underset{\rho_\textrm{i},\rho_\textrm{o}}{\text{maximize}}
& & I(Y;X), \\
& \text{subject to}
& &  \rho_\textrm{i}  \geq 0,\\
& & & \rho_\textrm{o}\geq \rho_\textrm{i},\\
& & & r_M^2\leq PAPR_0,
\end{aligned}
\label{eq:EqOpt2}
\end{equation}
where the input distribution for $X$ is now given by $x_m$ in \eqref{eq:Eqxm}, $r_m$ in \eqref{eq:Eqrm}, and where each constellation point have the probability $p_m=1/M$. After closer scrutiny, it is seen that the optimization problem in \eqref{eq:EqOpt2} can not easily be solved analytically. It is straightforward to further simplify the expressions (and the optimization) above by letting $\rho_\textrm{i} = 0$.

To simplify the expression for $r_m$ further, we may chose a form that converges to GB- and disc-GAM for $S=0$ and $S=\infty$, respectively. First let $\rho_\textrm{i}=0$. Inspecting \eqref{eq:Eqrm}, and knowing that $\rho_\textrm{o}$ diminishes with increasing SNR, we propose the simple, yet well-performing, form
 \begin{align}
r_m
&=\sqrt{\frac
{\ln
\left(1-\frac{m}{S+M}\right)}
{ \frac{1}{M}\sum_{m'=1}^M \ln
\left(1-\frac{m'}{S+M}\right)}}.
\label{eq:EqrmSNR}
\end{align}
When $S\rightarrow 0$, \eqref{eq:EqrmSNR} (essentially) converges to the GB-GAM's \eqref{eq:Eqgb} (by letting $m=\{0,1,\ldots,M-1\}$), whereas when $S\rightarrow \infty$, \eqref{eq:EqrmSNR}  converges to the disc-GAM's \eqref{eq:Eqdisc}.

\section{Numerical Results and Discussion}

\begin{figure}[tp!]
 \centering
 \vspace{-.4 cm}
 \includegraphics[width=9cm]{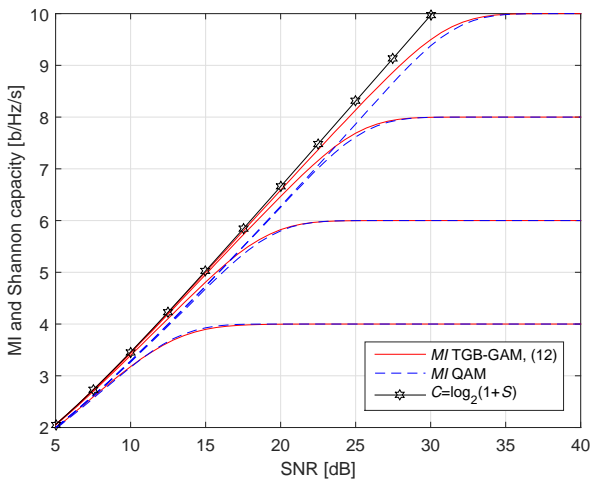}
 %\vspace{-5.6cm}
 \caption{MI of TGB-GAM \eqref{eq:EqrmSNR}, QAM, and the AWGN capacity,  for $H=\{4,6,8,10\}$.}
 \label{fig:TGBGAMQAM}
 \vspace{-.3cm}
\end{figure}

\begin{figure}[tp!]
 \centering
 \vspace{-.4 cm}
 \includegraphics[width=9cm]{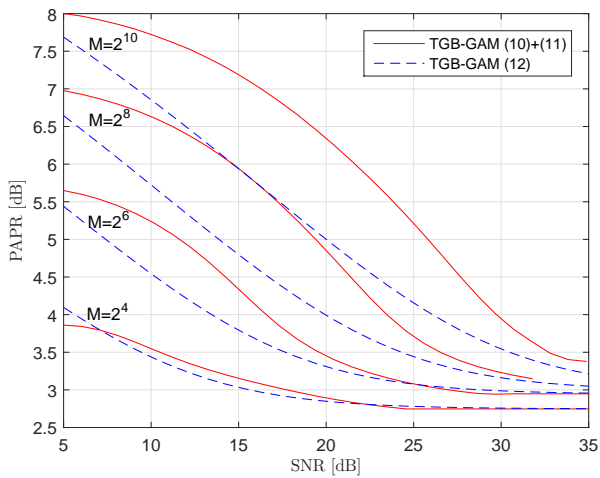}
 %\vspace{-5.6cm}
 \caption{PAPR vs. SNR of TGB-GAM \eqref{eq:Eqrm}+\eqref{eq:EqOpt2}, and TGB-GAM \eqref{eq:EqrmSNR}, for $H=\{4,6,8,10\}$.}
 \label{fig:TGBGAM_PAPR}
 \vspace{-.3cm}
\end{figure}

In the following, the MI vs. SNR performance for coded modulation (CM) is Monte Carlo simulated, whereas the optimized MI performance is numerically computed.
In Fig.~\ref{fig:ConstG2}\subref{fig:ConstG2-b}-\subref{fig:ConstG2-d}, we plot the signal constellations of optimized TGB-GAM \eqref{eq:Eqrm}+\eqref{eq:EqOpt2}, for  $\rho_\textrm{i}=0$, $M=256$, and $S= \{10, 22.5, 35\}$ dB. We note that the signal constellation change in shape from an approximate discrete complex Gaussian like design, over an intermediate design, to a disc-shaped design.
In Fig.~\ref{fig:MagnG2}\subref{fig:MagnG2-b}-\subref{fig:MagnG2-d}, we plot the magnitude distributions for the TGB-GAM together with GB-GAM and disc-GAM for the same SNRs as in Fig.~\ref{fig:ConstG2}\subref{fig:ConstG2-b}-\subref{fig:ConstG2-d}. We observe how the signal constellation magnitudes have nearly the same distribution as GB-GAM for low SNRs, and nearly the same distribution as disc-GAM for high SNR. We also note that the truncated Gaussian design overcomes the oscillatory problem seen in \cite{Larsson17b} for the GB-GAM G2 scheme at low SNRs. For the G2 scheme, a polynomial described the growth of $r_m$, and its coefficients where optimized to maximize the MI.
In Fig. \ref{fig:TGBGAMQAM}, we illustrate the MI-performance for the SNR-dependent TGB-GAM \eqref{eq:EqrmSNR} together with QAM and the AWGN capacity. It is observed that the MI performance of TGB-GAM is generally better than QAM, and approaches the AWGN capacity as SNR decreases. For $M=16$, TGB-GAM and QAM performs about the same.
%%%%%%%%%%%%%%%%%
%%%%%%%%%%%%%%%%%%
%%%%%%%%%%%%%%%%%
\begin{figure}[tp!]
 \centering
 \vspace{-.4 cm}
 \includegraphics[width=9cm]{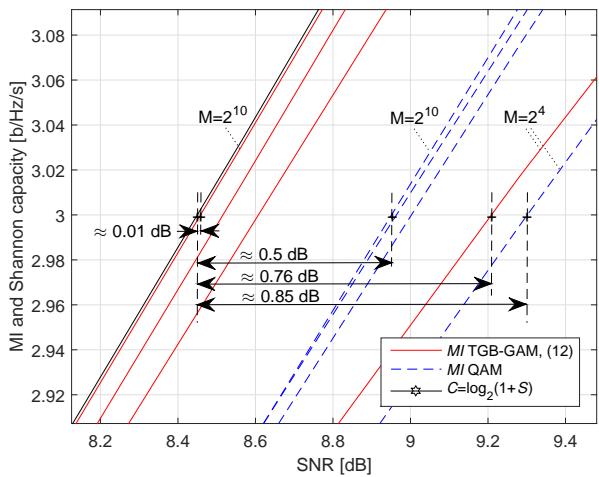}
 %\vspace{-5.6cm}
 \caption{MI of TGB-GAM \eqref{eq:EqrmSNR}, QAM, and the AWGN capacity, where $C\approx 3$ [b/Hz/s].}
 \label{fig:TGBGAMQAMCompare}
 \vspace{-.3cm}
\end{figure}
%%%%%%%%%%%%%%%%%%
%%%%%%%%%%%%%%%%%
\begin{figure}[tp!]
 \centering
 \vspace{-0.25cm}
 \includegraphics[width=9cm]{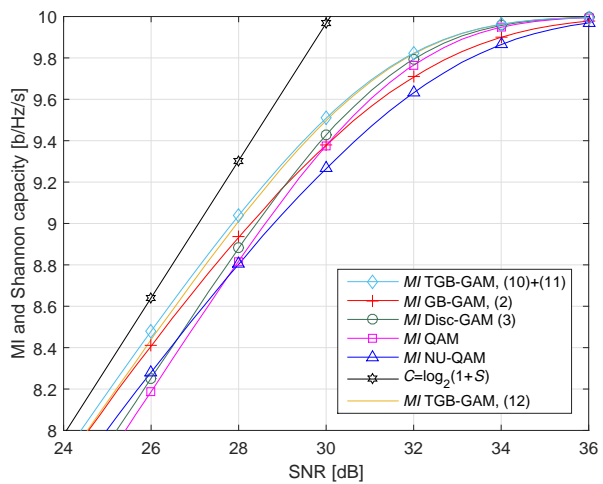}
 %\vspace{-5.6cm}
 \caption{MI of TGB-GAM \eqref{eq:Eqrm}+\eqref{eq:EqOpt2}, GB-GAM (2), disc-GAM (3), QAM, NU-QAM, and TGB-GAM \eqref{eq:EqrmSNR} with $M=2^{10}$, and the AWGN capacity.}
 \label{fig:GAM_TGB}
 \vspace{-0.3cm}
\end{figure}
%%%%%%%%%%%%%%%%%%
A possible misconception could be that since 16-QAM and 16-GAM have almost the same MI vs. SNR performance, there is no point in favoring GAM over QAM for relatively small constellation sizes, e.g. $M=16$. Such idea would be based on a certain design paradigm. Let's consider an example. Say that a rate $R=3$ [b/Hz/s] is desired. Then, the idea could be to use a constellation size $M=2^4$ and a channel code rate $r=3/4$. Zooming in around $C=3$ [b/Hz/s], Fig.~ \ref{fig:TGBGAMQAMCompare} shows that the SNR-gaps for 16-QAM and 16-TGB-GAM \eqref{eq:EqrmSNR} to the AWGN capacity at $C=3$ [b/Hz/s] are, respectively, $\approx 0.85$ dB and $\approx 0.76$ dB. (Note that the MI can be improved for GAM using \eqref{eq:Eqrm}+\eqref{eq:EqOpt2} or with optimization of $r_m$). However, this is not necessarily the best design approach, as the required SNR for a desired rate $R$ is unnecessarily high. Let's instead consider the MI-performance for very large $M$, say $M=2^{10}$. Then, the SNR-gaps from 1024-QAM and 1024-TGB-GAM \eqref{eq:EqrmSNR} to $C=3$ [b/Hz/s] are, respectively,  $\approx 0.5$ dB and $\approx 0.01$ dB. So a designer could on one hand choose 16-QAM with a $r=3/4$ code, or on another hand choose 1024-TGB-GAM with a rate $r=3/10$ code. The latter choice reduces the SNR-gap with $0.84$ dB. This has implication for modem design. Instead of implementing many different constellation sizes, one may implement just the largest desired constellation size, say $M=2^{10}$, and then adapt the code rate $r$ to achieve the desired rate $R$. Of course, this reasoning assumes that other costs, e.g. any additional energy consumption, are of less concern.
Another aspect speaking for GAM is that the PAPR performance. For example, in \cite{Larsson17a,Larsson17b}, we found that QAM asymptotically requires $\approx 1.96$ dB higher PAPR than for disc-GAM for the same average constellation point distances.
In Fig.~\ref{fig:GAM_TGB}, we compare the MI of TGB-GAM (both for \eqref{eq:Eqrm}+\eqref{eq:EqOpt2}, and \eqref{eq:EqrmSNR}), with GB-GAM, disc-GAM, QAM, and non-uniform QAM (NU-QAM) \cite[(1)-(2)]{SommerFett00}. As aimed for, TGB-GAM (both variants) performs better than all reference cases. NU-QAM performs quite poorly  when $MI\approx H$. Disc-GAM is, as expected due its disc-shape, about $0.2$ dB better than QAM.
In Fig.~ \ref{fig:TGBGAM_PAPR}, the PAPR vs. SNR of TGB-GAM \eqref{eq:Eqrm}+\eqref{eq:EqOpt2}, and TGB-GAM \eqref{eq:EqrmSNR}, for $H=\{4,6,8,10\}$ is plotted. The PAPR decreases for increasing SNR, as the constellation approaches the shape of disc-GAM. The PAPR is also higher for larger constellation sizes $M$.
%%%%%%%%%%%%%%%%%
%%%%%%%%%%%%%%%%%%

\section{Summary and conclusion}
In this work, we proposed a new GAM shaping design based on inverse sampling of (a double) truncated complex Gaussian pdf. The MI (and PAPR) performance inherently improved over the two special cases GB-GAM (good at low SNR) and disc-GAM (good at high SNR). A numerical optimization problem was formulated in only two (or optionally just one) optimization variables, and hence allows numerical optimization to be performed for large-$M$ constellations. A simplified, SNR-dependent, expression for the magnitude was proposed, and found to offer improved MI-performance over GB-GAM, disc-GAM, QAM, and NU-QAM. A capacity approaching, simplified implementation, modem design, with large size $M$ and adaptive low rate $r$, was discussed. It is hoped that GAM-based designs can find use in future wireless communication systems, such as next generation cellular, high-rate microwave-links, satellite, or optical fiber, systems.

\bibliographystyle{IEEEtran}
%\bibliography{./chapterbibl_42}


\end{document}